\documentclass[cameraready]{Interspeech}

\title{Towards Fine-grained Temporal Perception: Post-Training Large Audio-Language Models with Audio-Side Time Prompt}

\author[affiliation={1}]{Yanfeng}{Shi}
\author[affiliation={1}]{Pengfei}{Cai}
\author[affiliation={1}]{Jun}{Liu}
\author[affiliation={1}]{Qing}{Gu}
\author[affiliation={1}]{Nan}{Jiang}
\author[affiliation={1}]{Lirong}{Dai}
\author[affiliation={2}]{Ian}{McLoughlin}
\author[affiliation={1}, correspondingauthor]{Yan}{Song}

\address{
    $^1$ National Engineering Research Center of Speech and Language Information Processing, University
of Science and Technology of China, Hefei, China\\
    $^2$ ICT Cluster, Singapore Institute of Technology, Singapore
}

\email{yanf\_shi@mail.ustc.edu.cn, lrdai@ustc.edu.cn, ian.mcloughlin@singaporetech.edu.sg, songy@ustc.edu.cn}

\keywords{large audio-language model, fine-grained temporal perception, reinforcement learning}

\usepackage{comment}

\usepackage{booktabs}
\usepackage{multirow}
\usepackage[table]{xcolor}
\usepackage{tcolorbox}

\begin{document}

\maketitle

\begin{abstract}
    Large Audio-Language Models (LALMs) enable general audio understanding and demonstrate remarkable performance across various audio tasks.
    However, these models still face challenges in temporal perception (e.g., inferring event onset and offset), leading to limited utility in fine-grained scenarios.
    To address this issue, we propose Audio-Side Time Prompt and leverage Reinforcement Learning (RL) to develop the TimePro-RL framework for fine-grained temporal perception.
    Specifically, we encode timestamps as embeddings and interleave them within the audio feature sequence as temporal coordinates to prompt the model.
    Furthermore, we introduce RL following Supervised Fine-Tuning (SFT) to directly optimize temporal alignment performance.
    Experiments demonstrate that TimePro-RL achieves significant performance gains across a range of audio temporal tasks, such as audio grounding, sound event detection, and dense audio captioning, validating its robust effectiveness.
\end{abstract}

\section{Introduction}

Audio conveys a wealth of information, ranging from human speech to environmental events, and serves as a fundamental modality for perceiving the world\cite{audioset, panns}.
Large Audio-Language Models (LALMs) have significantly advanced general audio understanding by integrating the linguistic reasoning of Large Language Models (LLMs) with audio encoders~\cite{pengi, salmonn}.
These models have demonstrated strong versatility across a wide range of applications, such as acoustic scene classification, audio captioning and audio question answering~\cite{ltu, qwenaudio, audiopalm}.
Based on large-scale cross-modal pre-training, LALMs exhibit a remarkable ability to comprehend diverse acoustic content and interact with users through flexible natural language, providing a unified interface for audio-related tasks.

Despite these advancements, research indicates that current LALMs still exhibit shortcomings in fine-grained temporal understanding~\cite{timeaudio}.
In many practical applications, models are required not only to interpret acoustic content but also to capture the temporal structure or event boundaries within it\cite{sedtutorial, temporalqa}.
Although LALMs excel at semantic recognition, they often struggle to precisely infer the onset and offset timestamps of specific sound events.
This gap exposes a key limitation in fine-grained temporal perception, which is essential for temporally grounded tasks such as audio grounding\cite{ag} and sound event detection\cite{sed}.

Prior work has strengthened temporal perception through the construction of time-annotated datasets~\cite{audiotime}, yielding significant performance improvements.
Other approaches introduce specialized time tokens~\cite{groundedvideollm} to represent the concept of time, enhancing temporal alignment during generative inference.
While such advancements are promising, two aspects merit further attention: 
1) The audio input in LALMs lacks the explicit modeling of physical temporal cues, limiting the precise alignment between semantic content and its actual temporal coordinates.
2) Supervised Fine-Tuning (SFT) primarily focuses on semantic correctness, lacking optimization signals to address time-boundary prediction deviations.
Notably, Reinforcement Learning (RL) has shown potential in Video Temporal Grounding (VTG)~\cite{videochatr1, videotgr1}, where reward signals can be directly designed around temporal alignment metrics, offering valuable insights for analogous tasks in the audio domain.

Motivated by these observations, we propose the TimePro-RL framework to enhance the fine-grained temporal perception of LALMs via two primary modules: temporal information integration and training objective design.
Specifically, we introduce Audio-Side Time Prompt to inject temporal coordinates into audio features, then employ SFT to instruct the model in utilizing these coordinates.
Subsequently, we further develop RL post-training with an advantage-driven adaptive temporal reward to incorporate temporal alignment quality into the optimization objective.
Through the synergy of these strategies, TimePro-RL achieves significant performance gains across a range of audio temporal tasks.

\section{Method}

\begin{figure*}[!t]
    \centering
    \includegraphics[width=0.95\textwidth]{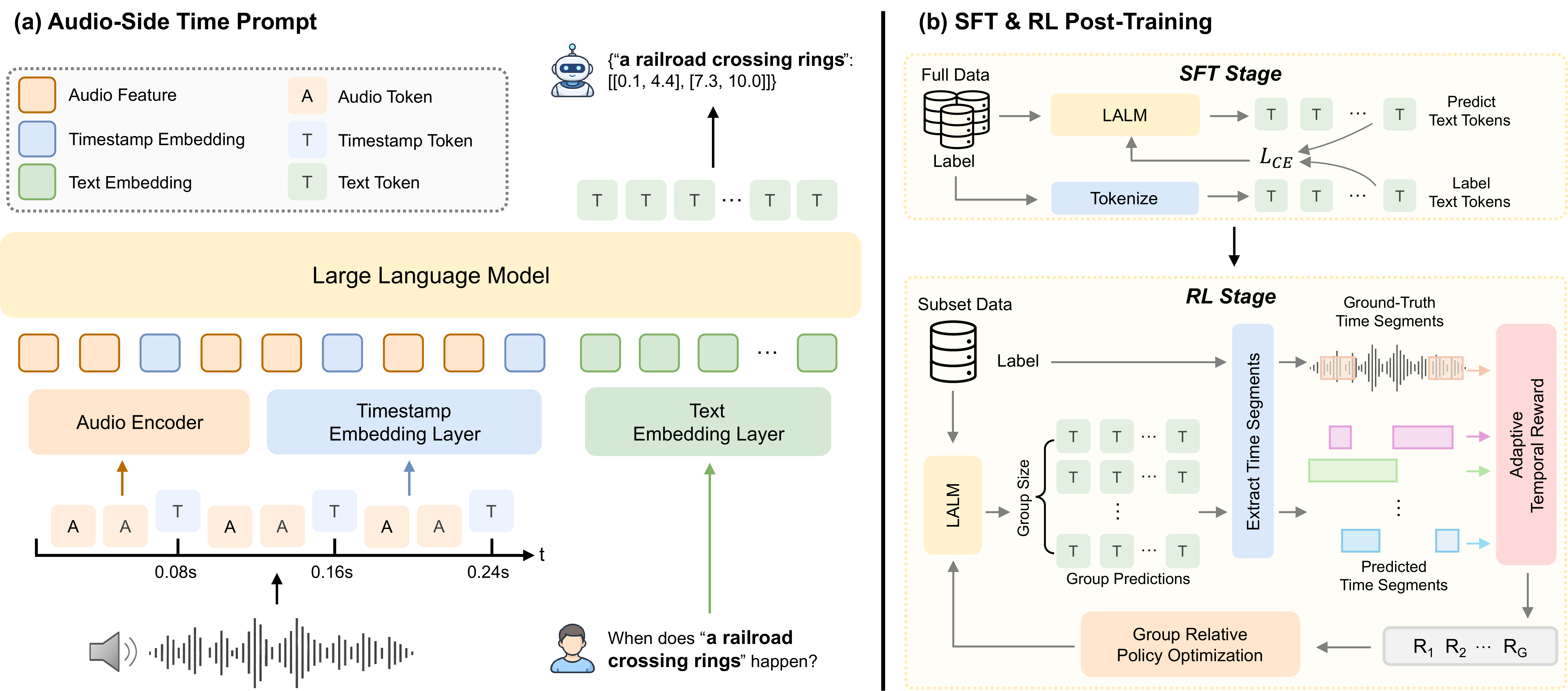}
    \caption{Overview of our TimePro-RL framework. Timestamp Embeddings are interleaved within audio features as time prompt, followed by SFT and RL post-training.}
    \label{fig:TimePro-RL}
\end{figure*}

In this section, we elaborate on the TimePro-RL framework, the overall schematic of which is illustrated in Figure~\ref{fig:TimePro-RL}.
We first present the infusion of temporal cues into LALM's audio input (Section~\ref{sec:ASTP}), and then describe the RL post-training paradigm tailored for audio temporal tasks, focusing on reward design (Section~\ref{sec:RL}).

\subsection{Audio-Side Time Prompt} \label{sec:ASTP}

Most LALMs predominantly rely on position embeddings, such as RoPE~\cite{rope}, to capture sequential structure.
While these models can be trained to extract temporal information from sequence inputs~\cite{timeunderstand}, directly inferring absolute timestamps remains challenging.
For VTG task, a previous study has overlaid frame indices onto video frames to provide perceptible time references, thereby enhancing temporal localization performance~\cite{numpro}.
The core of this approach lies in providing ``time prompt" at the input level, effectively reducing reasoning difficulty and mitigating hallucinations.

Inspired by this, we propose Audio-Side Time Prompt (ASTP), in which timestamps are encoded into embedding vectors and interleaved within the audio feature sequence to constitute LALM's audio input.
Specifically, we first extend the tokenizer with a set of Timestamp Tokens (e.g., \verb|<0.04>|), each corresponding to a specific time point in seconds.
During preprocessing, the audio input is partitioned into an audio token sequence based on its duration and the output frame rate of the audio encoder.
Since each position in this sequence maintains a fixed mapping to the timeline, Timestamp Tokens can be inserted into the sequence according to their corresponding time points, serving as explicit temporal coordinates to prompt the model.
The following shows an example of the processed input at the maximum time resolution:

\begin{tcolorbox}[
    colback = gray!6,
    colframe = black!60,      
    boxrule = 0.6pt,          
    rounded corners,          
    left = 2pt, right = 2pt,  
    top = 4pt, bottom = 4pt,
    fontupper = {\ttfamily\footnotesize\linespread{1.25}\selectfont}
]
<s><audio><AUDIO><0.04><AUDIO><0.08><AUDIO> \\
<0.12><AUDIO><0.16>$\cdots$</audio>When does "a railroad crossing rings" happen?</s>
\end{tcolorbox}

\noindent where \verb|<s>| and \verb|</s>| indicate the start and end of the sequence, while \verb|<audio>| and \verb|</audio>| demarcate the audio segment.
\verb|<AUDIO>| serves as a placeholder token, which is then replaced by the corresponding audio frame feature.
In this example, the audio encoder's output frame rate is \SI{25}{\hertz}, and tokens such as \verb|<0.04>| represent the inserted Timestamp Tokens.

Following this, Timestamp Tokens in the sequence are mapped to vector representations via the Timestamp Embedding Layer.
To ensure the stability of the embedding space, we adopt an initialization strategy based on semantic priors: each Timestamp Embedding is initialized as the mean of the subword embeddings derived from tokenizing its corresponding numerical string.
Taking \verb|<0.04>| as an example:

\begin{align}
  \mathbf{E}_{\langle 0.04 \rangle} &= \frac{1}{|\mathcal{T}(0.04)|} \sum_{u \in \mathcal{T}(0.04)} \mathbf{e}_{u} \label{eq:timestamp_init}
\end{align}

\noindent where $\mathbf{E}_{\langle 0.04 \rangle}$ indicates the embedding vector of Timestamp Token \verb|<0.04>|, $\mathbf{e}_{u}$ represents the pre-trained embedding of the token with ID $u$ in the vocabulary, and $\mathcal{T}(0.04)$ is the set of token IDs obtained by tokenizing the numerical string ``0.04".

Adding time prompt to LALM's input aligns well with its autoregressive architecture.
This enables the model to retrieve temporal information from neighboring Timestamp Embeddings when attending to event cues within audio frames.
Since the input structure is altered, SFT is required to guide the model in correctly understanding and utilizing this prompt.
The semantic initialization strategy facilitates this by transferring pre-trained knowledge from the original language model.
Moreover, all $\mathbf{E}_{\langle t \rangle}$ parameters are frozen during training to prevent semantic drift.

\subsection{RL for audio temporal tasks} \label{sec:RL}

\begin{table*}[!t]
    \centering
    \caption{Performance comparison of TimePro-RL and zero-shot / finetuned baseline models on audio temporal tasks. Bold indicates the best results in each category.}
    \label{tab:Results}
    \footnotesize
    \renewcommand\arraystretch{1.1} 
    \addtolength\tabcolsep{1pt} 
    \begin{tabular}{@{}lcccccccc@{}}
        \toprule
        \multirow{2}{*}{\textbf{Model}} & 
        \multirow{2}{*}{\textbf{Scale}} & 
        \multicolumn{4}{c}{\textbf{Audio Grounding}} & 
        \multicolumn{1}{c}{\textbf{Sound Event Detection}} & 
        \multicolumn{2}{c}{\textbf{Dense Audio Captioning}}  \\ 
        \cmidrule(lr){3-6} \cmidrule(lr){7-7} \cmidrule(lr){8-9}
        &                              & R@0.5 & R@0.7 & R@0.9 & mIoU & Eb-F1 & METEOR & Eb-F1 \\ 
        \midrule 
        \rowcolor[gray]{.95} \multicolumn{9}{c}{\textit{Zero-shot}} \\
        Qwen2-Audio & 7B               & 9.2 & 5.1 & 3.3 & 11.9 & 3.4 & 11.2 & 3.0 \\
        Qwen2.5-Omni  & 7B             & 25.4 & 17.4 & 10.6 & 27.7 & 13.7 & 10.5 & 10.4 \\
        \rowcolor[gray]{.95} \multicolumn{9}{c}{\textit{Finetuned}} \\
        Audio-Flaming2 & 3B            & 37.0 & 27.6 & 19.0 & 43.3 & 8.9  & 25.7 & 12.7 \\
        TimeAudio & 7B                 & 75.7 & 61.2 & 36.5 & 57.8 & $-$  & 20.4 & 37.4 \\
        Qwen2-Audio & 7B               & 74.8 & 57.9 & 34.6 & 69.6 & 49.8 & 32.2 & 35.0 \\
        Qwen2.5-Omni & 7B              & 74.0 & 59.8 & 34.1 & 69.9 & 48.9 & 31.3 & 35.2 \\
        Kimi-Audio & 7B                & 76.1 & 60.0 & 34.5 & 70.6 & 50.9 & 31.2 & 32.7 \\
        \midrule 
        \rowcolor[gray]{.95} \multicolumn{9}{c}{\textit{Post-Trained with TimePro-RL (\textbf{Ours})}} \\
        Qwen2-Audio & 7B               & 78.8 & 64.0 & 38.1 & 72.9 & \textbf{58.4} & \textbf{35.3} & 39.8 \\
        Qwen2.5-Omni & 7B              & \textbf{80.1} & \textbf{66.3} & \textbf{39.8} & \textbf{74.4} & 57.6 & 33.9 & \textbf{40.7} \\
        \bottomrule
    \end{tabular}
\end{table*}

While LALM’s temporal perception can be manifested in its ability to localize sound events, the standard SFT objective remains misaligned.
Consider a ground-truth segment [\SI{5.0}{\second}, \SI{6.0}{\second}], a reasonable prediction of [\SI{4.9}{\second}, \SI{5.9}{\second}] would be heavily penalized by token-level cross-entropy loss, which risks overfitting and impairs generalization capability~\cite{timer1}.
RL offers a viable path to address this problem by optimizing for evaluation metrics, and has been successfully applied in generation tasks like summarization by directly utilizing ROUGE or CIDEr as rewards~\cite{rougerl, ciderrl}.
However, the potential of RL for audio temporal tasks has yet to be extensively explored.

For further enhancing fine-grained temporal perception of LALMs, we extend the training process following SFT by adopting Group Relative Policy Optimization (GRPO)~\cite{grpo}, which computes relative advantages through group sampling.
To align the training objective with temporal alignment performance, we utilize the Event-based F1 score (Eb-F1)~\cite{ebf1}, an established metric in sound event detection, as the main reward ($\mathbf{r}_{\text{main}}$).
However, we observe in our experiments that since Eb-F1 is a threshold-based discrete metric, the limited group size in GRPO may result in identical rewards among sampled predictions, which leads to advantage degeneration and diminishes data efficiency.

To address this, we incorporate a continuous auxiliary reward ($\mathbf{r}_{\text{aux}}$), such as the mean Intersection over Union (mIoU), to provide smoother optimization signals.
We develop an advantage-driven adaptive temporal reward mechanism:

\begin{align}
    \mathbf{R} = 
    \begin{cases} 
    \mathbf{r}_{\text{main}} \odot \mathbf{r}_{\text{aux}}, & \text{if } \text{Var}(\mathbf{r}_{\text{main}}) < \epsilon \\ 
    \mathbf{r}_{\text{main}}, & \text{otherwise}
    \end{cases}
    \label{eq:adapt_reward}
\end{align}

\noindent where $\mathbf{R}$ indicates the group reward vector for advantage computation, $\text{Var}(\cdot)$ represents the variance operator, and $\epsilon$ is the variance threshold.
If $\mathbf{r}_{\text{main}}$ lacks discriminability among predictions in a group, the algorithm adopts the element-wise product of $\mathbf{r}_{\text{main}}$ and $\mathbf{r}_{\text{aux}}$ as the fused reward.
This strategy leverages the smoothness of $\mathbf{r}_{\text{aux}}$ to recover the advantage signal while utilizing $\mathbf{r}_{\text{main}}$ as a weight to regulate optimization intensity.
By dynamically adjusting the reward calculation based on real-time data conditions during training, this mechanism improves data efficiency while maintaining high temporal alignment quality.

\section{Experimental setup}

\subsection{Tasks and datasets}

We conduct experiments across three representative audio temporal tasks:

\textbf{Audio Grounding (AG)}~\cite{ag}. 
The audio grounding task is defined as localizing a specific sound event within an audio clip based on a descriptive natural language query. 
For this task, we utilize the FTAR dataset~\cite{timeaudio} and require the model to output the corresponding timestamps in the \texttt{\{"query": [onset, offset]\}} format. 
Performance is measured via Intersection over Union (IoU) metrics, where we report recall values at various thresholds including R@0.5, R@0.7, and R@0.9 alongside the mean IoU (mIoU).

\textbf{Sound Event Detection (SED)}~\cite{sed}. 
Sound event detection involves identifying event categories from a predefined set along with their respective occurrence periods. 
We conduct this task on the DESED dataset~\cite{desed}, with model outputs formatted as \texttt{\{"event": [onset, offset]\}}. 
The evaluation metric of this task leverages the Eb-F1 score~\cite{ebf1}, and we set the boundary tolerance to \SI{0.2}{\second}.

\textbf{Dense Audio Captioning (DAC)}~\cite{timeaudio}. 
For dense audio captioning, the model should generate descriptions for the sound events in the audio clip paired with their associated timestamps. 
This task is constructed using the FTAR dataset~\cite{timeaudio}, and outputs follow the format: \texttt{onset-offset, description}. 
To measure performance, we employ METEOR~\cite{meteor} to assess the linguistic quality of the generated captions and Eb-F1 to evaluate the accuracy of temporal localization.

The sample sizes for the training and test sets of each task are detailed in Table~\ref{tab:data}.

\begin{table}[!ht]
    \centering
    \caption{Summary of data statistics across the three tasks.}
    \footnotesize
    \renewcommand\arraystretch{1.2}
    \addtolength\tabcolsep{2pt}
    \begin{tabular}{lcc}
    \toprule
    \textbf{Task} & \textbf{Training Size} & \textbf{Test Size} \\ 
    \midrule
    Audio Grounding        & 61,862 & 483 \\
    Sound Event Detection  & 15,041 & 1,153 \\
    Dense Audio Captioning & 92,443 & 741 \\ 
    \bottomrule
    \end{tabular}
    \label{tab:data}
\end{table}

\subsection{Implementation details}

\begin{table*}[!t]
    \centering
    \caption{Ablation study of different components. ASTP denotes Audio-Side Time Prompt; ``random init" refers to random initialization of Timestamp Embeddings; RL(Eb-F1) indicates using only Eb-F1 as the reward.}
    \label{tab:Ablation}
    \footnotesize
    \renewcommand\arraystretch{1.2} 
    \addtolength\tabcolsep{1pt} 
    \begin{tabular}{@{}lccccccc@{}}
        \toprule
        \textbf{Method} & 
        \multicolumn{4}{c}{\textbf{Audio Grounding}} & 
        \multicolumn{1}{c}{\textbf{Sound Event Detection}} & 
        \multicolumn{2}{c}{\textbf{Dense Audio Captioning}} \\ 
        \cmidrule(lr){2-5} \cmidrule(lr){6-6} \cmidrule(lr){7-8}
        \textbf{(Qwen2.5-Omni)}   & R@0.5 & R@0.7 & R@0.9 & mIoU & Eb-F1 & METEOR & Eb-F1 \\ 
        \midrule 
        SFT Baseline              & 74.0 & 59.8 & 34.1 & 69.9 & 48.9 & 31.3 & 35.2 \\
        \midrule 
        w/ ASTP (random init)     & 73.2 & 57.2 & 32.8 & 68.8 & 46.0 & 31.4 & 33.3 \\
        w/ ASTP                   & 77.6 & 61.7 & 35.8 & 71.7 & 50.1 & 32.6 & 37.0 \\   
        w/ ASTP + RL (Eb-F1)      & 77.8 & 63.1 & 38.9 & 72.7 & 56.9 & 31.6 & 38.1 \\   
        w/ ASTP + RL              & 80.1 & 66.3 & 39.8 & 74.4 & 57.6 & 33.9 & 40.7 \\
        \bottomrule
    \end{tabular}
\end{table*}

To evaluate the effectiveness of the TimePro-RL framework, we utilize Qwen2-Audio~\cite{qwen2audio} and Qwen2.5-Omni~\cite{qwen2.5omni} as base models. 
Both of them integrate Whisper~\cite{whisper} as the audio encoder with an output frame rate of \SI{25}{\hertz}.
To support Audio-Side Time Prompt at the maximum time resolution, we expand the tokenizer with 750 Timestamp Tokens, covering from \SIrange{0}{30}{\second} with a stride of \SI{0.04}{\second}.
We conduct parameter-efficient fine-tuning using LoRA~\cite{lora} with $r$~=~8 and $\alpha$~=~32. 
The overall post-training pipeline consists of two stages:
1) \textbf{SFT}: The models are trained on the full dataset for 3 epochs with a learning rate of $1\times10^{-5}$.
2) \textbf{RL}: Following SFT, we take GRPO for only a single epoch using a subset of 10,200 samples. 
The group size is set to 4, and the learning rate is $1\times10^{-6}$. 
In the adaptive temporal reward mechanism, we employ Eb-F1 as $\mathbf{r}_{\text{main}}$ across all three tasks. 
For $\mathbf{r}_{\text{aux}}$, we utilize mIoU for AG and SED, while METEOR for DAC. 
The variance threshold $\epsilon$ is specified as $1\times10^{-6}$.

\section{Results}

In this section, we first evaluate TimePro-RL against baselines under zero-shot and finetuned settings.
Then, we conduct ablation studies to analyze the contributions of components in TimePro-RL.

\subsection{Main results}

As shown in Table~\ref{tab:Results}, we first evaluate Qwen2-Audio and Qwen2.5-Omni under zero-shot conditions, where their performance on high-precision metrics (e.g., R@0.9 and Eb-F1) is notably constrained, revealing the limitations of existing general-purpose LALMs in fine-grained temporal perception.
Subsequently, we compare models post-trained via TimePro-RL framework with several prominent LALMs adapted by SFT on the same dataset, including Qwen2-Audio, Qwen2.5-Omni, Audio-Flamingo2~\cite{audioflamingo2} and Kimi-Audio~\cite{kimiaudio}.
For TimeAudio, which is only trained on the FTAR dataset, we adopt the results reported in its original publication~\cite{timeaudio}.
The results show that TimePro-RL consistently outperforms these baselines across multiple evaluation metrics, and crucially, it demonstrates a strong competitive edge in high-precision localization.
For instance, in AG task, Qwen2.5-Omni improves from 34.1 during the SFT stage to 39.8 on R@0.9.
Similarly, in DAC task, its Eb-F1 score rises from 35.2 to 40.7, while Qwen2-Audio also achieves a 4.8-point improvement.
These gains validate the effectiveness of TimePro-RL in enhancing the fine-grained temporal perception of LALMs.

\subsection{Ablation study}

We conduct a series of ablation experiments based on Qwen2.5-Omni to systematically investigate the contributions of each component in TimePro-RL framework, with the results summarized in Table~\ref{tab:Ablation}.
Compared to the SFT baseline, using random initialization in Audio-Side Time Prompt leads to performance regressions across most metrics, such as a 2.9-point drop in SED Eb-F1, because randomly initialized Timestamp Embeddings introduce extraneous noise into the audio feature sequence.
In contrast, the semantic initialization strategy proposed in Section~\ref{sec:ASTP} allows the model to correctly interpret and leverage Timestamp Embeddings as temporal coordinates, offering performance gains including a 1.7-point increase in AG R@0.9.
This underscores the critical importance of semantic initialization for Audio-Side Time Prompt.

Furthermore, our results demonstrate that even a modest volume of RL training significantly boosts model performance, with AG R@0.9 increasing from 35.8 to 39.8 compared to the SFT (w/ ASTP) stage.
In terms of the reward design, the adaptive temporal reward mechanism in Section~\ref{sec:RL} provides more balanced overall gains compared to the Eb-F1 only configuration.
While the latter causes an optimization imbalance where the METEOR score for DAC task declines from the 32.6 achieved in the SFT (w/ ASTP) stage to 31.6, our adaptive approach recovers this score to 33.9 and pushes the final Eb-F1 to 40.7, facilitating more comprehensive data utilization.

\begin{figure}[!t]
    \centering
    \includegraphics[width=\columnwidth]{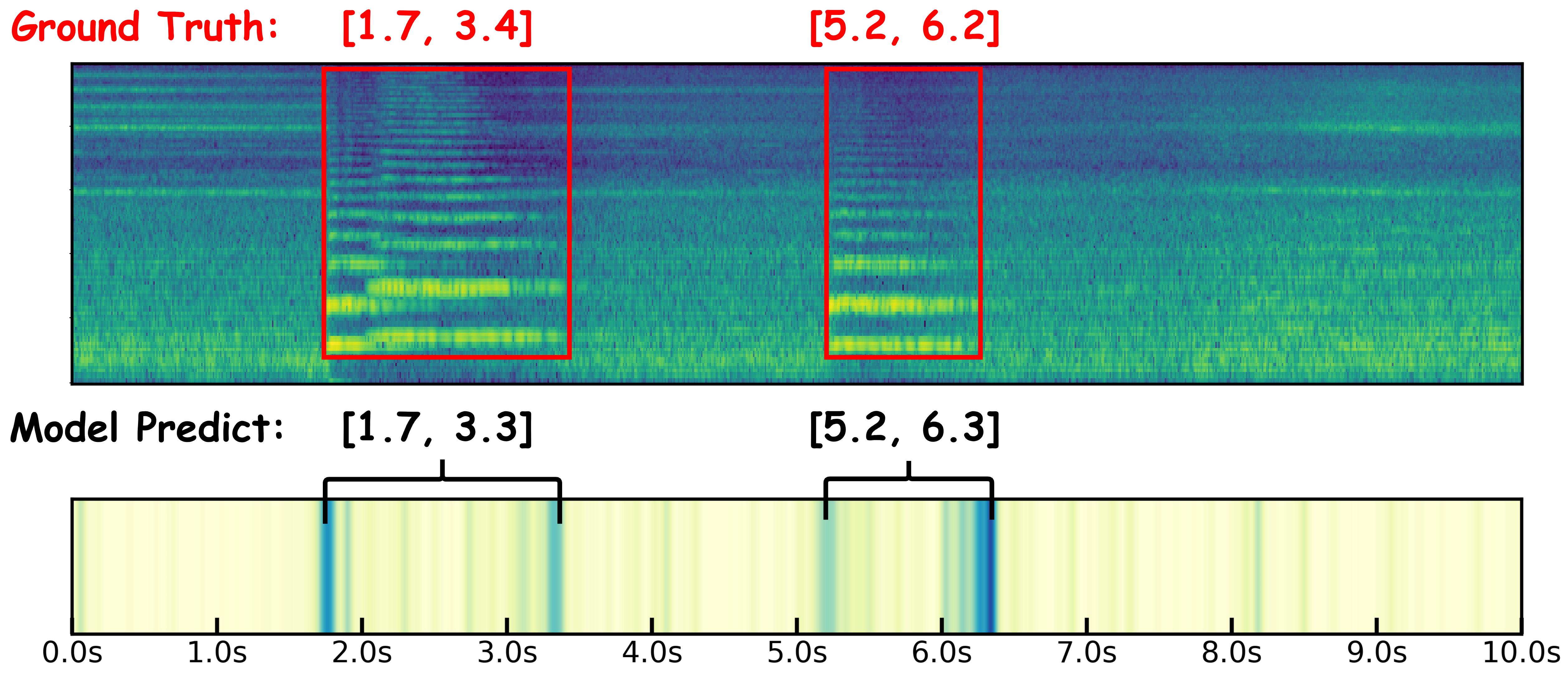}
    \caption{Visualization of attention weights on Timestamp Embeddings. For the audio grounding query ``a train horn honking", the mel-spectrogram (top) is aligned with the chronologically arranged attention weights assigned to each Timestamp Embedding (bottom).}
    \label{fig:Attention}
\end{figure}

Figure \ref{fig:Attention} provides a visual analysis of the attention weights assigned to Timestamp Embeddings to further elucidate the internal mechanism of our framework.
we extract the attention weights of the generate tokens toward each Timestamp Embedding from the model's final layer and plot them along the timeline.
The attention map reveals that the model's focus exhibits high-intensity activations that precisely align with the onset and offset boundaries of the sound events marked in the mel-spectrogram.
This sharp concentration of attention on the start and end coordinates suggests that the model effectively leverages Timestamp Embeddings to capture precise temporal cues for sound events.
Such clear alignment provides intuitive evidence for the interpretability of Audio-Side Time Prompt, confirming that it successfully guides the model in perceiving and utilizing fine-grained temporal information integrated in the audio feature sequence.

\section{Conclusion}

In this paper, we propose TimePro-RL, a framework to enhance the fine-grained temporal perception of LALMs. 
TimePro-RL interleaves Timestamp Embeddings into audio feature sequence to provide temporal cues, and introduces RL post-training with an adaptive temporal reward designed for temporal alignment to further strengthen temporal capabilities. 
Experiments show that TimePro-RL achieves significant performance gains across a range of audio temporal tasks, validating the effectiveness of the synergy between Audio-Side Time Prompt and RL post-training. 
Future research will explore the application of our method in complex reasoning scenarios, such as Chain-of-Thought (CoT), where fine-grained temporal cues serve as critical intermediate evidence.

\clearpage

\bibliographystyle{IEEEtran}
\bibliography{mybib}

\end{document}